# Understanding and measuring software engineer behavior: What can we learn from the behavioral sciences?

Allysson Allex Araújo[1], Marcos Kalinowski[2], and Daniel Graziotin[3]

[1]Federal University of Cariri (UFCA)
Center for Science and Technology
Juazeiro do Norte, CE – Brazil

[2]Pontifical Catholic University of Rio de Janeiro
(PUC-Rio) Department of Informatics
Rio de Janeiro, RJ – Brazil

[3]University of Hohenheim
Institute of Information
Systems Stuttgart – Germany

`allysson.araujo@ufca.edu.br`, `kalinowski@inf.puc-rio.br`,
`graziotin@uni-hohenheim.de`

***Abstract.*** *This paper explores the intricate challenge of understanding and measuring software engineer behavior. More specifically, we revolve around a central question: How can we enhance our understanding of software engineer behavior? Grounded in the nuanced complexities addressed within Behavioral Software Engineering (BSE), we advocate for holistic methods that integrate quantitative measures, such as psychometric instruments, and qualitative data from diverse sources. Furthermore, we delve into the relevance of this challenge within national and international contexts, highlighting the increasing interest in understanding software engineer behavior. Real-world initiatives and academic endeavors are also examined to underscore the potential for advancing this research agenda and, consequently, refining software engineering practices based on behavioral aspects. Lastly, this paper addresses different ways to evaluate the progress of this challenge by leveraging methodological skills derived from behavioral sciences, ultimately contributing to a deeper understanding of software engineer behavior and software engineering practices.*

## 1. What is the biggest proposed challenge?

The examination of human behavior has garnered considerable attention across different academic domains. Consequently, behavioral sciences were born to study human behavior [McConnell 1974, Skinner 1965]. This interest arises from the complex task of comprehending human behavior, which results from a complex interplay among actions, cognition, and emotion [Carter 2017]. Psychology, for instance, has long studied behaviorism, which posits that all behaviors are learned through environmental interactions [Watson 2017]. Building upon this perspective, Lenberg *et al.* (2015) proposed the Behavioral Software Engineering (BSE) field as a specialized one concerned with human aspects of Software Engineering (SE). In particular, they argue that delineating a distinct SE field centered on realistic human attributes is relevant for enhancing comprehension and refining practices within software development processes.

The interest in the behavioral facet of SE has spotlighted a considerable challenge: *How can we enhance our understanding of software engineer behavior?* A possible avenue toward addressing this question involves the thorough collection of both quantitative and qualitative empirical data. For instance, quantitative measures facilitated by psychometric instruments have emerged as valuable tools in ensuring the systematic development and interpretation of psychological tests [Graziotin et al. 2022]. Psychometrics is the field concerned with the development of measurement instruments and the assessment of whether these instruments are reliable and valid forms of measurement [Furr 2021]. Also, qualitative data could be gleaned from diverse sources such as in-depth interviews, focus group sessions, and behavioral observations [Lenberg et al. 2015].

In addressing why this challenge is relevant, we align with Feldt *et al.* (2008) and Graziotin *et al.* (2021) regarding the importance of systematically integrating precise measurements to advance more rigorous scientific theories and yield substantiated results within SE. By employing robust quantitative and qualitative methods, researchers can capture and analyze the intricacies of software engineer behavior, thus paving the way for a deeper understanding of the underlying mechanisms driving software development processes [Lenberg et al. 2014]. In terms of software engineer behavior, we may exemplify works covering happiness [Graziotin and Fagerholm 2019], impostor phenomenon [Guenes et al. 2023], burnout [Tulili et al. 2023], emotions [Kurian and Thomas 2023], ethics [Johnson and Menzies 2023], etc. For industrial applications, the strategic deployment of behavioral analysis holds strong potential in organizations [Wilder et al. 2009]. For example, how do software engineers behave and cognitively approach the task of designing a software architecture? What methods can be used to measure the behavior, cognitive processes, and emotional states involved? By gaining insights into the software developer's behavior with precise measurements, software organizations can gain guidelines into the behavioral dynamics of SE processes, facilitating data-driven decision-making and optimizing SE efficiency. In other words, understanding software engineer behavior may ultimately enhance operational efficiency, foster a culture of evidence-based practice, and drive continuous improvement within SE practices [Petre et al. 2020].

## 2. What is the specific context related to it and its relevance in the national and/or international context of social, human, and economic aspects of software?

Despite considerable advancements, the efficacy of psychometric measurements in empirical SE research has been undermined by a pervasive misinterpretation of associated constructs and their methodologies [Graziotin et al. 2022, Felipe et al. 2023]. Graziotin *et al.* (2015), for example, have observed that SE scholars tend to confuse affect-related psychological constructs such as emotions and moods with related, yet different, constructs such as motivation, commitment, and well-being. This misalignment becomes particularly evident when validated psychological tests are adapted by the SE community, often resulting in modifications to test items that compromise the tests' psychometric reliability and validity [Gren and Goldman 2016, Gren 2018, Felipe et al. 2023]. Consequently, while psychometrics are valuable, their universal acceptance remains contentious within psychology. Critically, psychometric-based assessments may overlook information in direct interactions with individuals, notably qualitative data, thus warranting a nuanced evaluation of measurement strategies [Graziotin et al. 2022, Schoenherr and Hamstra 2016].

Hence, quantitative methods are just one aspect of a complex issue, and a mixed-method approach that includes qualitative studies could be necessary to deeply understand software engineer behavior [Lenberg et al. 2014]. In this sense, Lenberg *et al.* (2017) advocate for integrating diverse qualitative methods drawn from behavioral sciences, such as grounded theory, interpretive analysis, ethnography, phenomenology, narrative analysis, and discourse analysis. In other words, one can argue about the valuable opportunity of addressing other epistemological and ontological positions to understand the phenomenon under investigation [Ogundare 2017]. However, it is also essential to acknowledge that establishing criteria for qualitative research poses notable challenges, particularly given the scarcity of standards developed within the BSE domain. In navigating these challenges, SE researchers must exercise prudence and reflexivity in selecting and applying qualitative methods [Lenberg et al. 2023]. This diverse toolkit offered by qualitative psychology holds promise for SE empirical investigations by ensuring methodological rigor and validity of research outcomes [Molléri et al. 2018].

Thus, while BSE studies may not yet dominate the mainstream discourse in SE research, the field remains nascent with a growing body of literature and evolving knowledge. Indeed, BSE presents promising opportunities for specialized academic venues domestically and internationally. Noteworthy examples include the Workshop on Social, Human, and Economic Aspects of Software (WASHES) in Brazil, the International Conference on Cooperative and Human Aspects of Software Engineering (CHASE), and the Software Engineering in Society track of the International Conference on Software Engineering (ICSE-SEIS). These serve as qualified forums for scholarly-industry exchange, leveraging interdisciplinary dialogues and advancing research agendas in BSE. At the national level, initiatives focused on understanding and assessing software engineer behavior hold the potential to shape workforce development strategies, policies, and efforts to bolster the software industry. Similarly, international collaborations and knowledge-sharing endeavors in this domain offer avenues for cultivating best practices, standards, and policies that promote ethical conduct, diversity, and inclusivity in software development [Carver et al. 2021]. In summary, given the global nature of the software industry and its interconnected workforce, initiatives that promote cross-cultural understanding, collaboration, and innovation could drive positive outcomes.

## 3. What real initiatives is it related to?

Different real-world initiatives are underway to deepen our understanding of BSE, although not all exclusively focus on this aspect. Prominent examples within the software industry include the 'Stack Overflow Developers Survey', the 'SAP Developer Insights Survey', and the 'JetBrains State of Developer Ecosystem'. These surveys gather data on various aspects of software developer behavior, including technological preferences, job satisfaction, and career aspirations. Through analysis of survey findings, researchers gain insights into the motivational factors, challenges, and emerging trends that influence the behavior of software engineers in industrial settings. Another noteworthy endeavor is the DevOps Research and Assessment (DORA) Program, backed by Google. As articulated on DORA's official website, their "research team applies behavioral science methodology to uncover the predictive pathways which connect ways of working, via software delivery performance, to organizational goals and individual well-being". Furthermore, several organizations and consulting companies, particularly those specializing in Developer Experience (DevEx), agile transformation, and change management, are increasing their focus on BSE to improve their triage efforts and improvement programs. As an illustrative example, Microsoft has launched the Developer Experience Lab (DevExLab) whose objective is "to discover, improve, and amplify developer work and well-being".

In addition, various academic initiatives are advancing BSE. Chalmers University's Department of Computer Science, through Robert Feldt and Richard Torkar, focuses on BSE research. The Chair of Software Engineering at the Technical University of Munich, Heilbronn, directed by Stefan Wagner, and the Department of Information Systems and Digital Technologies at the University of Hohenheim, under Daniel Graziotin, apply empirical and behavioral methods to study software engineering and digital transformation. Also, Maria Teresa Baldassarre leads BSE research at the University of Bari in the Software Engineering Research Laboratory. In Brazil, Marcos Kalinowski has taken the lead in conducting research on BSE under the ExACTa R&D initiative at the Pontifical Catholic University of Rio de Janeiro. Kiev Gama has also explored this field at the Informatics Center of the Federal University of Pernambuco.

These real-world initiatives emphasize the interest and investment in comprehending software engineer behavior. Through the utilization of varied data sources, methods, and collaborative efforts, both researchers and practitioners stand to advance our understanding of behavioral issues and refine SE practices across different domains.

## 4. Ways to evaluate the progress of the proposed challenge

In addition to the implications of what can be done to precisely understand software engineer behavior (in theory and practice), SE researchers stand to gain advantages by harnessing the methodological skills derived from the behavioral sciences to conduct relevant BSE studies. According to Gren (2018), it is imperative to increase the prevalence of studies dedicated to introducing, validating, and utilizing psychometric instruments in BSE. Graziotin *et al.* (2021), for example, provided a valuable discussion on psychometric theory tailored for SE researchers, offering guidelines for both utilizing existing instruments and developing new ones. Their comprehensive review of psychology literature, framed within the Standards for Educational and Psychological Testing, outlined important activities in operationalizing new psychological constructs. These activities encompass item pooling, item review, pilot testing, item analysis, factor analysis, and assessment of statistical properties such as reliability, validity, and fairness, including considerations for test bias. More recently, Felipe *et al.* (2023) have conducted a systematic mapping on psychometric instruments for assessing personality within SE.

In addition, Green and Goldman (2016) advocate for adopting underutilized statistical methods in human factors research within SE, such as Test-Retest, Cronbach's $\alpha$, and exploratory factor analysis, all of which are pertinent to psychometric assessment. Green (2018) also proposed a psychological test theory framework for characterizing validity and reliability in BSE research, reinforcing the necessity for maintaining fair psychometric properties. Collectively, these works emphasize the importance of integrating robust psychometric principles into behavioral research methodologies within SE, thereby facilitating rigorous and reliable empirical investigations in psychometric-based SE analysis.

On the other hand, according to Lenberg *et al.* (2017a), seminal criteria outlined by Lincoln and Guba (1985), Maxwell (1992), and Sandelowski (1986) have profoundly influenced the evaluation of qualitative research methodology. Lincoln and Guba (1985) proposed five criteria for naturalistic inquiries: credibility, transferability, dependability, confirmability, and authenticity. In turn, Maxwell (1992) further emphasized the importance of integrity and criticality, while Sandelowski (1986) advocated for creativity and artfulness in qualitative inquiry. Lenberg *et al.* (2017) concluded that future qualitative studies would benefit from adopting a broader set of qualitative research methods, emphasizing reflexivity, and employing qualitative guidelines and quality criteria.

Therefore, establishing robust theoretical and methodological underpinnings should constitute a foundational step in designing measurement approaches to properly understand software engineer behavior. Within BSE research, particularly in exploring psychological constructs, there remains a notable gap in adopting rigorous and validated research artifacts [Graziotin et al. 2022, Guimarães et al. 2021]. Hence, we need to cultivate awareness and appreciation among SE researchers for theories and tools from established behavioral sciences, including on the perspective of SE education [Araújo et al. 2024]. Finally, to assess the progress in addressing the identified challenge, we believe the WASHES community would embrace methodological approaches that meet the established criteria we discussed earlier. It is also important to link research findings with practical applications, and vice versa, in the industry. Additionally, regular assessments (surveys, longitudinal studies, etc.) and baseline metrics could be used to track progress over time. By incorporating lessons from psychology and related behavioral disciplines, we may improve the methodological foundations of BSE and, hopefully, gain a deep understanding of the behavioral dynamics inherent to the SE context.